\documentclass[sn-standardnature]{sn-jnl}% Standard Nature Portfolio Reference Style
\usepackage{lmodern}   % scalable CM lookalike
\usepackage{appendix} % To handle appendices
\usepackage{caption}
\usepackage{graphicx} % Required for inserting images
\usepackage{amsmath}
\usepackage[utf8]{inputenc}
\usepackage[T1]{fontenc}
\usepackage{amsmath}
\usepackage{siunitx}
\usepackage{xcolor}
\usepackage{booktabs}
\usepackage{physics}
\usepackage{caption}
\usepackage{amsfonts}
\usepackage{graphicx} % Required for inserting images
\usepackage{amsmath}
\usepackage{float}
\usepackage{xcolor}
\usepackage{siunitx}
\usepackage{booktabs}
\usepackage{physics}
\usepackage{caption}
\usepackage{amsfonts}

\newcommand{\rr}[0]{\textbf{r}}
\newcommand{\kk}[0]{\textbf{k}}
\newcommand{\bq}[0]{\textbf{q}}
\newcommand{\QQ}[0]{\textbf{Q}}

\newcommand{\eV}[0]{\text{eV}}

\newcommand{\mos}[0]{$\text{MoS}_2$ }

\theoremstyle{thmstyleone}%
\theoremstyle{thmstyletwo}%

\theoremstyle{thmstylethree}%

\begin{document}

\title[Real-Space Imaging of Moiré-Confined Excitons in Twisted Bilayer MoS$_2$]{Real-Space Imaging of Moiré-Confined Excitons in Twisted Bilayer MoS$_2$}

\author[1]{\fnm{Laurens J.M.} \sur{Westenberg}}
\author[2]{\fnm{Lumen} \sur{Eek}}
\author[1]{\fnm{Jort D.} \sur{Verbakel}}
\author[1]{\fnm{Kevin} \sur{Vonk}}
\author[1]{\fnm{Stijn J.H.} \sur{Borggreve}}

\author[3]{\fnm{Kenji} \sur{Watanabe}}
\author[4]{\fnm{Takashi} \sur{Taniguchi}}

\author[1]{\fnm{Paul} \sur{de Boeij}}
\author[5]{\fnm{Rodrigo} \sur{Arouca}}
\author[2]{\fnm{Cristiane} \sur{Morais Smith}}
\author*[1]{\fnm{Pantelis} \sur{Bampoulis}}\email{p.bampoulis@utwente.nl}

\affil[1]{\orgdiv{Physics of Interfaces and Nanomaterials, MESA+ Institute for Nanotechnology}, \orgname{University of Twente}, \orgaddress{\street{Drienerlolaan 5}, \city{Enschede}, \postcode{7522 NB}, \country{the Netherlands}}}

\affil[2]{\orgdiv{Institute for Theoretical Physics}, \orgname{Utrecht University}, \orgaddress{\street{Princetonplein 5}, \city{Utrecht}, \postcode{3584 CC}, \country{the Netherlands}}}

\affil[3]{\orgdiv{Research Center for Electronic and Optical Materials}, \orgname{National Institute for Materials Science}, \orgaddress{\street{1-1 Namiki}, \city{Tsukuba}, \postcode{305-0044}, \country{Japan}}}

\affil[4]{\orgdiv{Research Center for Materials Nanoarchitectonics}, \orgname{National Institute for Materials Science}, \orgaddress{\street{1-1 Namiki}, \city{Tsukuba}, \postcode{305-0044}, \country{Japan}}}

\affil[5]{\orgdiv{Centro Brasileiro de Pesquisas Físicas}, \orgaddress{\street{R. Dr. Xavier Sigaud, 150}, \city{Rio de Janeiro - RJ}, \postcode{22290-180}, \country{Brazil}}}



\abstract{

\textbf{Twisted two-dimensional semiconductors generate a moiré landscape that confines excitons (bound electron-hole pairs) into programmable lattices, offering routes to efficient light sources, sensing, and room-temperature information processing. However, direct real-space imaging of confined excitonic species within a moiré unit cell remains challenging; existing claims are inferred from spatially averaged far-field signals that are intrinsically insufficient to resolve nanometre-scale variations. Here, we imaged excitons across the moiré of a 2$^{\circ}$ twisted bilayer MoS$_2$ with nanometre resolution using room-temperature photocurrent atomic force microscopy. We directly resolved site-selective confinement: direct and indirect excitons localize at different stacking registries of the moiré, with contrast governed by alignment between site-selective generation and confinement minima. A Wannier-based moiré-exciton model reproduces the measured energies and the moiré-induced localization of the exciton wavefunction. These species-specific, unit-cell-resolved measurements constrain microscopic models of moiré excitons, provide benchmarks for excitonic order, and establish a device-compatible route to engineering excitonic lattices in van der Waals heterostructures.}
\\

}

 % adjust the value to taste: -1.0cm to -2.0cm usually works
\maketitle

Twisting two atomically thin van der Waals layers turns a uniform crystal into a nanoscale moiré potential that reshapes its electronic and optical properties \cite{cao2018unconventional,yu2017moire,baek2020highly,guo2025superconductivity}. This matters especially in transition-metal dichalcogenides (TMDs), where photoexcitation creates tightly bound excitons with valley-dependent selection rules that remain stable at room temperature \cite{splendiani2010emerging,mueller2018exciton,huang2022excitons,du2023moire,de2025advanced}. In this setting, the moiré potential modulates local band edges and optical matrix elements on the nanometer scale, tuning exciton energies and real-space wavefunctions \cite{carr2020electronic,yu2017moire,baek2020highly,huang2022excitons}. Under suitable conditions it confines excitons into periodic lattices, enabling programmable emitters, electrical excitonic sensing, analogue processing, and access to topological bands and Bose-Hubbard physics in a solid-state platform \cite{wu2017topological,yu2017moire,jin2019observation,tran2019evidence,carr2020electronic,baek2020highly}. 
However, a species-resolved real-space picture of where different excitons localize within a single moiré cell remains elusive \cite{yu2017moire,huang2022excitons, du2023moire, de2025advanced}. This is because far-field optics average over thousands of moiré cells, and current real-space probes either require cryogenic temperatures, unit-cell averaging, or image only one excitation \cite{seyler2019signatures,andersen2021excitons,karni2022structure,tizei2015exciton,susarla2022hyperspectral,naik2022intralayer,li2024imaging}.

Here, we use room-temperature photocurrent atomic force microscopy (pc-AFM), in a device-compatible geometry, to map the excitonic response of a 2$^{\circ}$ twisted bilayer MoS$_2$ with sub-nanometre spatial and high spectral resolution. Real-space maps reveal site-selective confinement across the moiré: direct (A, B) and indirect (e.g., $\Gamma$–$K$, $i$) excitons occupy distinct stacking registries, with contrast set by the alignment between site-selective generation and the confinement minima. These species-resolved, unit-cell-level measurements provide unambiguous conclusions on exciton confinement informing theoretical models and bringing a new level of control to light-matter interactions at the nanoscale \cite{jauregui2019electrical,yu2017moire,du2023moire, kennes2021moire,ciarrocchi2022excitonic}.

\begin{figure}
    \centering
    \includegraphics[width=\linewidth]{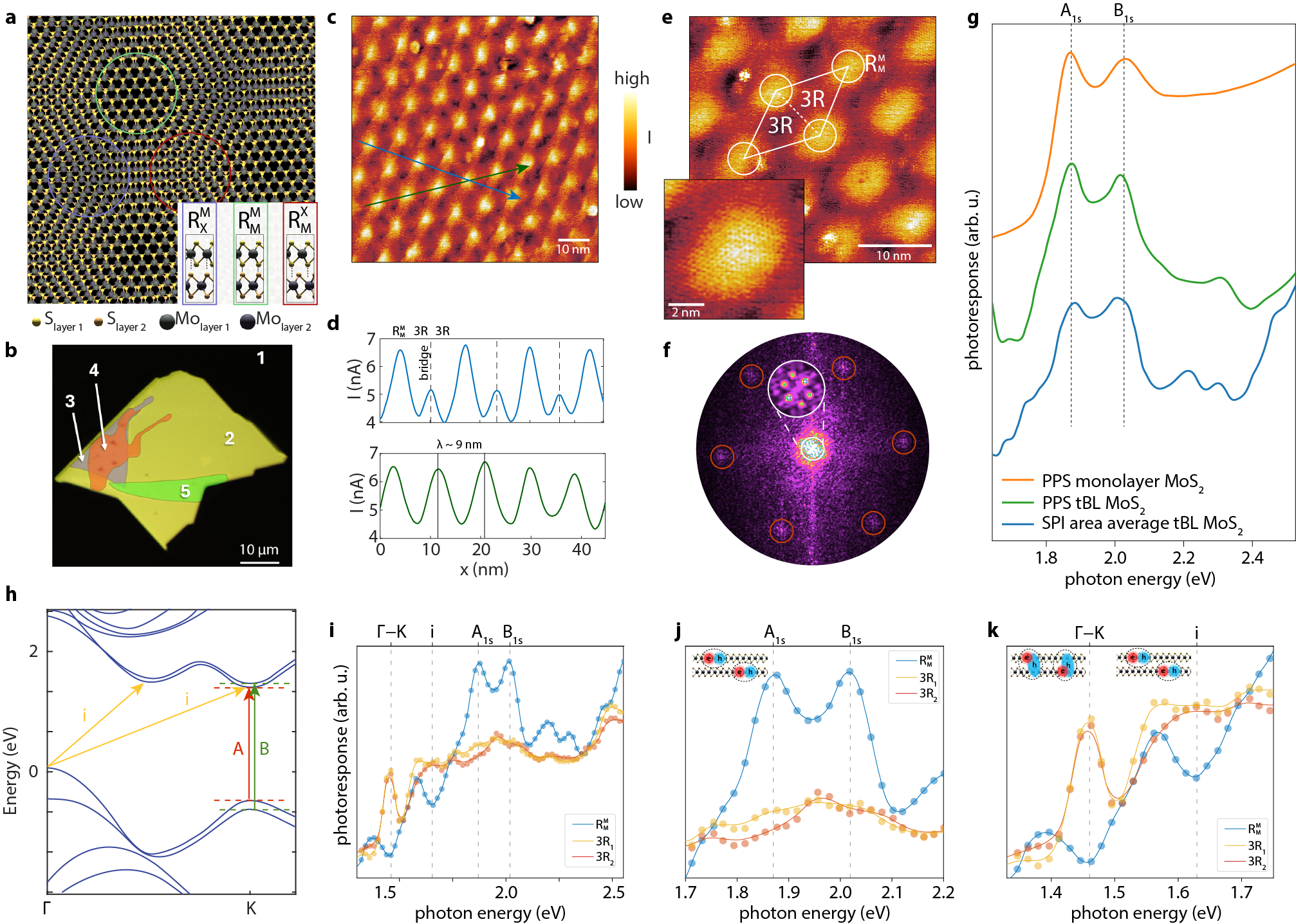}
    \caption{\textbf{Moiré lattice and photocurrent spectroscopy in twisted bilayer MoS\(_2\).} \textbf{a,} The rigid structural model of twisted bilayer MoS\(_2\) labeling the three high-symmetry registries ($R_{M}^{M}$ and $R_{M}^{X}$ and $R_{X}^{M}$, marked with the corresponding colors). \textbf{b,} Optical microscopy image of our \(\theta=2^{\circ}\) MoS\(_2\) bilayer on hBN on mica sample. The different regions on the sample are marked: (1) mica, (2) hBN, (3) bottom MoS\(_2\), (4) top MoS\(_2\), (5) graphene contact. \textbf{c,} Large-area c-AFM image resolving a uniform hexagonal moiré lattice. \textbf{d,} Cross-sections acquired along the two lines in (\textbf{c}) showing the moiré periodicity \(\lambda_{\mathrm{m}}\approx 9~\mathrm{nm}\) and the size of the different domains, with the bridges extended similarly to the nodes. \textbf{e,} Atomic-resolution c-AFM zoom showing the moiré lattice and the $1\times1$ MoS$_2$ lattice (see also inset for a zoom-in). The three high-symmetry sites ($R_{M}^{M}$ and two 3R sites) and the intersections (bridges) are marked in the image. \textbf{f,} FFT of the moiré marking all the relevant spots, consistent with \(\alpha\approx 0.32~\mathrm{nm}\), \(\lambda\approx 9~\mathrm{nm}\) and \(\theta\approx2.0^{\circ}\). \textbf{g,} PPS of monolayer MoS\(_2\) (orange) showing room-temperature A$_{1s}$ and B$_{1s}$ excitons at \(1.87\) and \(2.03~\mathrm{eV}\) (splitting \(\sim160~\mathrm{meV}\)). PPS (green) and SPI (blue) of the \(2^{\circ}\) twisted bilayer exhibit A$_{1s}$/B$_{1s}$ peaks at \(1.87\) and \(2.02~\mathrm{eV}\) and additional structure near \(2.2~\mathrm{eV}\) consistent with higher-order Rydberg states. \textbf{h,} Band-structure schematic indicating direct excitons (A$_{1s}$,B$_{1s}$) and indirect intralayer excitons (\(\Gamma - K,\Lambda\)). \textbf{i,} Registry-resolved SPI \(\mathrm{PR}(E)\) spectra averaged over pixels assigned to each stacking. \textbf{j,} Zoom-in on the A$_{1s}$/B$_{1s}$ energy region showing strong enhancement at $R_{M}^{M}$ and suppression at 3R sites. \textbf{k,} Zoom-in on the indirect excitons showing the opposite site selectivity: enhanced \(\Gamma- K\) (\(\sim1.46~\mathrm{eV}\)) and \(i\) (\(1.6\text{-}1.7~\mathrm{eV}\)) features at 3R and strong suppression at $R_{M}^{M}$. All spectra are recorded at room temperature.}

    \label{fig:fig1}
\end{figure}

When two TMD layers are twisted by a small angle $\theta$, a moiré superlattice emerges (Fig.~\ref{fig:fig1}a). For near $0^\circ$ twist angles, the local stacking is 3R type, and the three high-symmetry registries are the $R_{M}^{M}$, $R_{M}^{X}$, and $R_{X}^{M}$ stacking sites, as noted in Fig.~\ref{fig:fig1}a. Fig.~\ref{fig:fig1}b shows our $2^{\circ}$ twisted MoS$_2$ bilayer on hexagonal boron nitride (hBN) supported by mica sample. The sample is contacted with graphene to improve the contact resistance. Large-area conductive-AFM (c-AFM) maps resolve a uniform hexagonal moiré lattice (Fig.~\ref{fig:fig1}c) with a period of $\lambda_{\mathrm{m}}\approx 9~\mathrm{nm}$, as seen from the line profiles in Fig.~\ref{fig:fig1}d. Atomic-resolution images and fast Fourier transform (FFT) (Figs.~\ref{fig:fig1}e,f) confirm $\lambda_{\mathrm{m}}\approx 9~\mathrm{nm}$, i.e., consistent with the intended  $\theta\approx 2^{\circ}$ twist angle. The moiré pattern (Figs.~\ref{fig:fig1}c,d) features broad, nearly straight lines (‘bridges’) that trace the boundaries between $R_{M}^{X}$ and $R_{X}^{M}$ stacking sites, with their intersections (‘nodes’) corresponding to $R_{M}^{M}$ stacking sites, that manifest as the brightest c-AFM features in Figs.~\ref{fig:fig1}c,e. The three high-symmetry registries ($R_{M}^{M}$, $R_{M}^{X}$ and $R_{X}^{M}$) occupy nearly equal area fractions (within $\sim$5\%), indicating minimal relaxation~\cite{weston2020atomic,rosenberger2020twist}, in line with the rigid-limit structural model in Fig.~\ref{fig:fig1}a. As the $R_{M}^{X}$ and $R_{X}^{M}$ sites are electronically nearly degenerate, combined with the indistinguishability between them using c-AFM, we henceforth refer to these as `3R' sites.

To probe the registry-dependent response of the moiré lattice, we employ pc-AFM, which combines c-AFM with monochromatic illumination to obtain nanoscale, spectrally resolved maps of the photocurrent. A schematic of our set-up is shown in Extended Data Fig.~\ref{fig:extdatafig1}a. The sample is back-illuminated by a fiber-coupled monochromator fed by a xenon lamp ($280-1100$~nm). The optical spot diameter exceeds the device to ensure spatially uniform excitation of the sample. A small constant DC bias voltage \(V\) ($-1$ V) is applied between tip and sample while we record the tip current at position \(\mathbf r\) and photon energy \(E\) under illumination, \(I_{\ell}(\mathbf r,E,V)\), and in the dark, \(I_{d}(\mathbf r,V)\). We define the photocurrent \(I_{\mathrm{ph}}(\mathbf r,E,V)=I_{\ell}-I_{d}\) and the dimensionless photoresponse
\[
\mathrm{PR}(\mathbf r,E,V)=\frac{I_{\mathrm{ph}}(\mathbf r,E,V)}{\,q\,P(E)/E}\,,
\]
where \(q\) is the elementary charge and \(P(E)\) is the optical power at the sample plane. This normalization removes spectral variations of the lamp/monochromator power and dark current spatial variations, thus yielding a power-independent metric. We operate pc-AFM in two complementary modes that emphasize different aspects of the three-dimensional data set \(\mathrm{PR}(\mathbf r,E,V)\): photoresponse point spectroscopy (PPS) and scanning photoresponse imaging (SPI). In PPS, the tip is placed on a specific location and \(\mathrm{PR}(E)\) is recorded as \(E\) is swept at fixed \(V\). A single sweep typically takes \(2-3\) minutes. Because tip drift over that time is comparable to a significant fraction of the moiré period (a few nanometers), the PPS spectra should be interpreted with reduced spatial specificity. In SPI, the tip raster-scans the surface at a fixed photon energy \(E\) to acquire a \(\mathrm{PR}(\mathbf r)\) image. This is repeated for successive energies. The resulting image stack is then registered (sub-pixel cross-correlation, see Methods), which cancels drift and returns a \(\mathrm{PR}(E)\) spectrum at every pixel with subnanometer-scale spatial fidelity. The trade-off is that each SPI frame requires \(\sim 30~\mathrm{s}\), so SPI uses coarser energy steps and exhibits higher noise as continuous scanning increases susceptibility to tip changes. Despite these trade-offs, PPS and SPI are complementary modes, with PPS providing finer spectral resolution albeit with lower spatial confidence, while SPI offering robust spatial contrast with coarser spectral resolution. 

%Nevertheless, as we show below, PPS and SPI yield consistent spectral features and relative contrasts across registries, providing confidence in both the spectral and spatial aspects of the measurements.

The PPS spectrum of monolayer MoS$_2$ (Fig.~\ref{fig:fig1}g) exhibits the hallmark A$_{1s}$ ($1.87~\mathrm{eV}$) and B$_{1s}$ ($2.03~\mathrm{eV}$) excitons, separated by $\approx 160~\mathrm{meV}$, set primarily by the valence-band spin-orbit splitting (see band-structure schematic in Fig.~\ref{fig:fig1}h) \cite{vaquero2020excitons,wu2015exciton}. Over the experimental ranges explored, the energy peak positions and integrated areas are not measurably affected by sample bias, tip material, or applied load, indicating that the tip does not introduce a discernible perturbation to the spectra (Extended Data Fig.~\ref{fig:extdatafig1}). Extended Data Figs.~\ref{fig:extdatafig1}b-g further documents the thickness-dependent redshift of the exciton energies, and includes analogous benchmark spectra for other TMDs. These measurements establish that pc-AFM yields quantitative, spectrally faithful PR($E$) under room temperature and N$_2$ environment. Indeed, the $2.0^{\circ}$ twisted MoS$_2$ bilayer on hBN/mica (Fig.~\ref{fig:fig1}g) shows PPS (green) and SPI (blue) spectra with well-resolved A$_{1s}$ and B$_{1s}$ peaks at $1.87~\mathrm{eV}$ and $2.02-2.03~\mathrm{eV}$. The A$_{1s}$ - B$_{1s}$ spacing and peak energies are consistent with the monolayer spectra, our theory calculations (see Methods) and with prior reports on 3R-type bilayers \cite{xiong2022twist,agunbiade2024transient,ouyang2025electrically}. Compared to the monolayer, in our twisted bilayer spectra we resolve additional structure near $2.2~\mathrm{eV}$, consistent with higher-order Rydberg excitons, see Methods and Refs. \cite{wu2015exciton, marino2018quantum, vaquero2020excitons}. This improved contrast may arise from enhanced screening of surface charges by the hBN substrate.

\begin{figure}
    \centering
    \includegraphics[width=\textwidth]{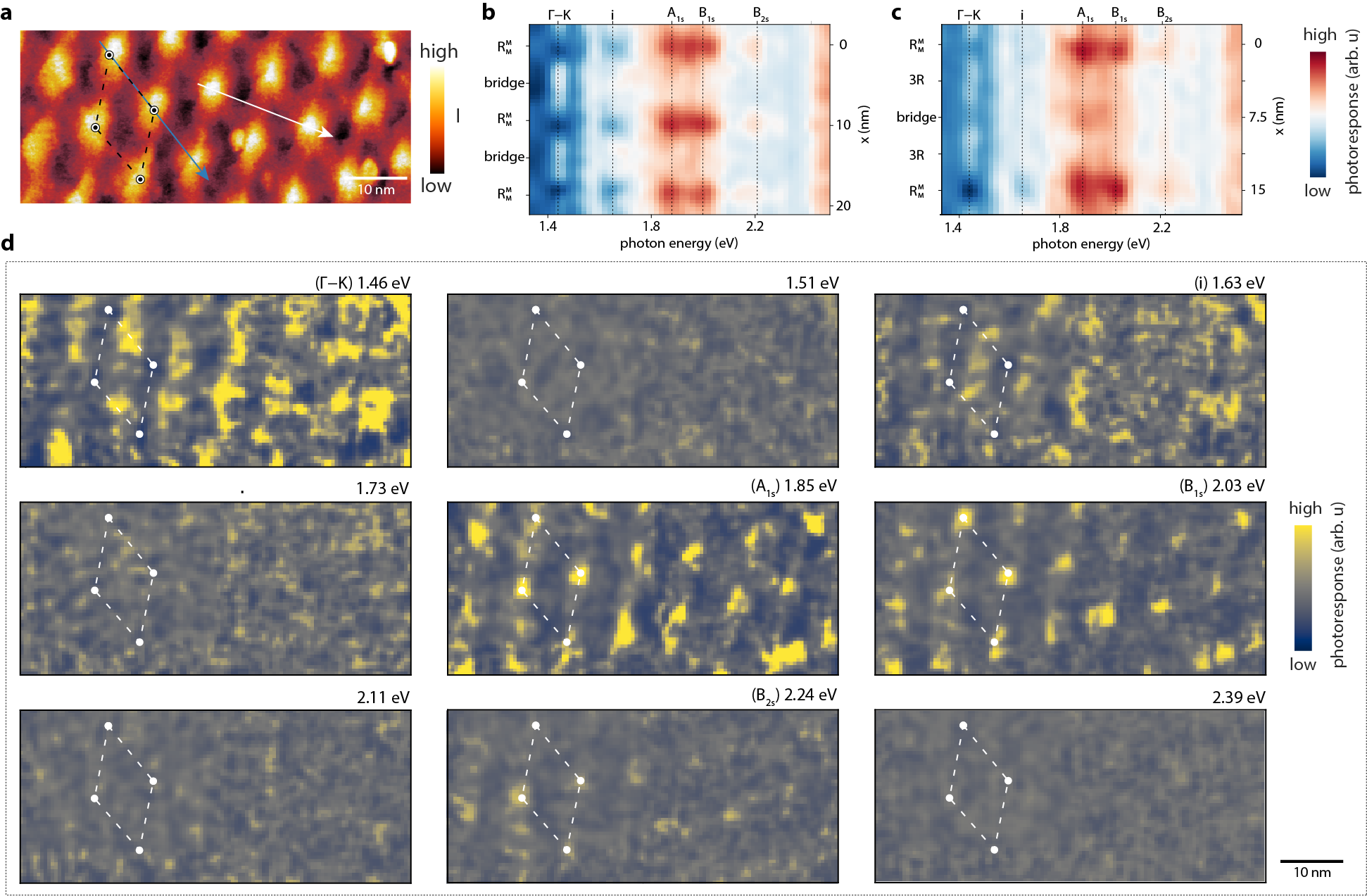}
    \caption{\textbf{Real-space, registry-selective exciton confinement in the moiré superlattice.} \textbf{a,} Dark current image resolving the moiré lattice, the arrows mark scan trajectories used for line spectroscopy. \textbf{b,c} \(\mathrm{PR}(E)\) linespectroscopies along the blue and white paths in \textbf{a,} showing periodic enhancement of the direct A$_{1s}$ and B$_{1s}$ excitons and their Rydberg states near \(2.2~\mathrm{eV}\) at $R_{M}^{M}$ nodes, and revealing features at \(\sim1.46~\mathrm{eV}\) (\(\Gamma- K\), interlayer) and \(1.6\text{-}1.7~\mathrm{eV}\) (\(i\), intralayer) that localize at 3R domains and anticorrelate with the A$_{1s}$/B$_{1s}$ contrast. \textbf{d,} Energy-selected PR maps off resonance as well as at A$_{1s}$, B$_{1s}$, \(\Gamma- K\), and \(i\) transitions, demonstrating registry-locked localization and small site-dependent energy shifts.}

    \label{fig:fig2}
\end{figure}

Exploiting the nanometer-scale spatial precision of SPI, we map how the moiré superlattice modulates the photoresponse. The dark-current image in Fig.~\ref{fig:fig1}e resolves the moiré lattice with atomic-scale detail and marks the three high-symmetry stacking sites ($R_{M}^{M}$ and two 3R sites). For each registry within the unit cell, we extract and average SPI spectra to obtain \(\mathrm{PR}(E)\) with registry specificity (full range in Fig.~\ref{fig:fig1}i, zoom-ins in the energy range $1.7 - 2.2$ eV in Fig.~\ref{fig:fig1}j and in the energy range $1.35 - 1.7$ eV in Fig.~\ref{fig:fig1}k). In addition, we acquire \(\mathrm{PR}(E)\) line spectroscopy along paths that traverse many moiré cells (arrows in Fig.~\ref{fig:fig2}a), shown in Figs.~\ref{fig:fig2}b,c. Both the registry-averaged spectra and the line scans reveal that the A$_{1s}$ ($1.87~\mathrm{eV}$) and B$_{1s}$ ($2.03~\mathrm{eV}$) excitons, and their higher-order features near \(2.2-2.4~\mathrm{eV}\), are strongly site dependent (Fig.~\ref{fig:fig1}j and Figs.~\ref{fig:fig2}b,c). They are pronounced at $R_{M}^{M}$ nodes and strongly suppressed at 3R sites, providing evidence for clear registry selectivity across the moiré unit cell. 

%This behaviour contrasts with rigid-limit expectations for twisted homobilayers, where a weak intralayer moiré potential would produce only a modest first-harmonic modulation of oscillator strength with little real-space contrast \cite{wu2017topological, huang2022excitons}, or with relaxed bilayers where the lowest energy states are predicted to be the 3R domains \cite{tilak2022moire}. Our data thus suggests that earlier works underestimated the intralayer moiré potential of rigid structures. 

In contrast, at the 3R stacking sites (Fig.~\ref{fig:fig1}k and Figs.~\ref{fig:fig2}b,c) we observe features near \(1.46~\mathrm{eV}\) and \(1.6-1.7~\mathrm{eV}\). These features are suppressed at $R_{M}^{M}$ sites. The \(\sim 1.46~\mathrm{eV}\) feature is consistent with an indirect interlayer \(\Gamma - K\) exciton \cite{liang2022optically,yang2024non,sung2020broken}. Its selectivity at 3R can be rationalized by asymmetric interlayer coupling under the applied out-of-plane electric field together with the intrinsic polar stacking of 3R domains, which favors interlayer dipoles. The \(1.6-1.7~\mathrm{eV}\) feature can be assigned to the indirect intralayer \(i\) exciton \cite{mak2010atomically, zhao2021strong}, arising from multiple transitions, like the \(\Gamma - K\) and \(\Gamma - \Lambda\) transitions (Fig.~\ref{fig:fig1}h). These experimentally observed peaks correspond to well-defined Wannier excitons, see theory section in Methods, reinforcing their identification as \(\Gamma - K\) and \(\Gamma - \Lambda\) transitions. Unlike photoluminescence, where indirect excitons are typically dark, photocurrent is sensitive to generation and extraction of carriers. Combined with moiré-assisted scattering and excitonic hybridization, this can relax selection rules, enabling detection of these excitonic channels \cite{murayama2005structure,yagodkin2023probing,quereda2017observation}. 

The registry dependence of the direct and indirect excitons is corroborated by energy-selected PR maps (Fig.~\ref{fig:fig2}d). Both the line scans and the maps (Figs.~\ref{fig:fig2}b-d) display clear, species-selective confinement of the excitonic photoresponse across the moiré, with an anticorrelation between the direct (A/B) and indirect (\(\Gamma- K\), \(i\)) channels. Moreover, further inspection of the line spectra in Figs.~\ref{fig:fig2}b,c reveal a confinement length of about $\sim$~2~nm, matching our theory expectations (see Methods), as well as small site-to-site energy variations of the excitons. We attribute this to modest strain and registry differences. Taken together, these data provide real-space evidence that the moiré superlattice imprints a registry-dependent, species-selective excitonic response. For completeness, in Extended Data Figs.~\ref{fig:extdatafig2}a-e, we provide additional sets of data acquired at different locations, showing the same exciton response. 

\begin{figure}[H]
    \centering
    \includegraphics[width=0.5\linewidth]{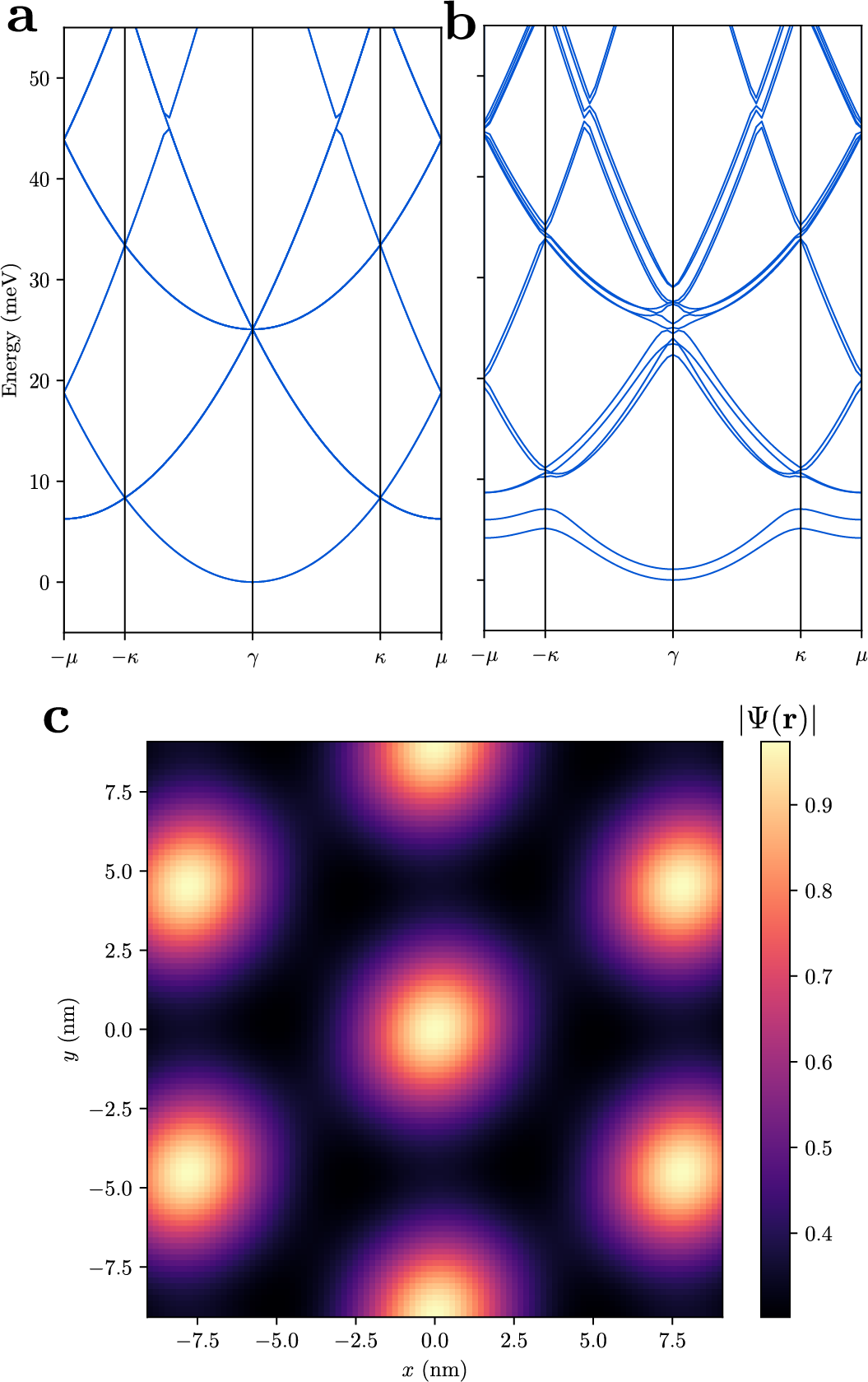}
    \caption{\textbf{Exciton bandstructure and wavefunction localization}. $K-K$ Exciton moiré band structure of twisted bilayer MoS$_2$. Band structure for \textbf{a,} uncoupled and \textbf{b,} coupled $2^\circ$ twisted MoS$_2$ bilayers. \textbf{c,} Excitonic wavefunction at the $\gamma$ point for the lowest-energy moiré band in \textbf{b}. The colormap indicates the exciton wavefunction magnitude.}
    \label{fig:MEX}
\end{figure}

Our local photoresponse measurements reflect four factors: (1) generation (set by local oscillator strength and band alignment), (2) transport within an effective length, (3) dissociation (governed by the tip field), and (4) extraction across the tip-MoS$_2$ Schottky barrier. Current-voltage characteristics under both polarities yield nearly equal Schottky barrier heights that are independent of stacking registry (see Methods and Extended Data Figs.~\ref{fig:extdatafig2}f-g), consistent with Fermi-level pinning at the tip-MoS$_2$ contact~\cite{sotthewes2019universal}. The nearly uniform barrier implies a uniform tip-induced field ($\sim 0.5-1~\mathrm{Vnm^{-1}}$), making local dissociation (under the tip) and extraction rates registry-independent~\cite{lee2019layer, quereda2021role, haastrup2016stark, kamban2019field}. What remains is where excitons are created and how far they move before dissociation. Generation is strongly site-selective because oscillator strength and detuning vary with stacking \cite{zhang2017interlayer, susarla2022hyperspectral}, so excitons are created preferentially at specific sites. If the transport length is comparable to the moiré period, diffusion would smear population into neighboring sites. However, our measurements show strong suppression at non-preferred sites with an exciton extent of $\sim 2~\mathrm{nm}$ and an off-site photoresponse that is negligible. Therefore, the observed spatial contrast reflects co-alignment of site-selective generation with ultrafast intra-cell relaxation to confinement minima.

To test this hypothesis, we modeled excitons in bilayer MoS$_2$ using a 2D screened interaction with parabolic bands for intra- and interlayer species, and incorporated interlayer tunneling at $K$ via a zone-folded exciton Hamiltonian (see Methods). With parameters appropriate to our stack (effective background $\kappa_0 \approx 2$), the calculations yield binding energies of the order of $\sim-300$~meV (intralayer) and $\sim -200$~meV (interlayer). For the direct $K-K$ excitons, this places the A$_{1s}$ and B$_{1s}$ near $1.85$ and $2.03$~eV, with the A$_{2s}$ at $2.05$~eV and the B$_{2s}$ at $2.2$~eV, consistent with the PPS/SPI spectra (see Extended Data Figs.~\ref{fig:binding}a-c). In addition, the model predicts interlayer \(\Gamma - K\) excitons with binding energies of $\sim200$~meV and transition energies around $1.45-1.5$~eV, as well as intralayer $\Gamma - \Lambda$ excitons with comparable binding energies and transition energies near $1.6-1.7$~eV, matching our experimental results. At $\theta\approx2^\circ$, diagonalization of the moiré exciton Hamiltonian produces hybridized minibands whose lowest bright state concentrates real-space weight at $R_{M}^{M}$ registries, Figs.~\ref{fig:MEX}a-c. In this framework, the large amplitude contrast is explained by interlayer hybridization and local symmetry strongly modulating the oscillator strength, while modest gap and binding-energy variations only weakly affect the exciton energies. The calculated exciton probability density $\lvert \Psi(\mathbf{r}) \rvert^2$ indeed concentrates at $R_{M}^{M}$ sites for the lowest bright band (Figs.~\ref{fig:MEX}c), in line with the experimental results. For indirect excitons, our Wannier-based calculations reproduce the transition energies observed in experiment. Their localization at 3R sites is qualitatively consistent with polarization-assisted interlayer coupling in polar 3R domains under an out-of-plane field \cite{tang2024interlayer, yang2024non, liang2022optically}. Overall, our theory accounts quantitatively for the direct exciton confinement and exciton energies, and provides a microscopic framework that also rationalizes the registry selectivity of the indirect species. 

Beyond establishing a decisive real-space picture of excitonic moiré lattices, our results demonstrate that distinct exciton species are confined at different registries, with direct and indirect excitons anticorrelated across the moiré unit cell. This species-resolved mapping clarifies the microscopic role of the moiré landscape and provides a benchmark for theories of excitonic order in twisted bilayers. At the same time, by introducing pc-AFM as a room-temperature, device-compatible, and spectrally accurate probe, we open a route to (i) test models of moiré excitons, topological excitons and correlated phases \cite{wu2017topological, xiong2023correlated}, (ii) engineer programmable excitonic lattices as a platform for Bose-Hubbard physics \cite{gotting2022moire, park2023dipole}, and (iii) realize nanoscale optoelectronic elements such as field-tunable emitters, valley filters, and quantum-light sources \cite{yu2017moire, ciarrocchi2019polarization, baek2020highly, du2023moire}. Because pc-AFM can be applied broadly across van der Waals heterostructures, its reach extends well beyond MoS$_2$ to designer moiré materials.

\section*{Methods}

\subsection*{Sample preparation} 
We assembled a twisted MoS$_2$ homobilayer on hBN supported by mica. Bulk MoS$_2$ (acquired from HQ graphene) and hBN crystals were exfoliated into few-layer flakes using the Scotch tape method. These flakes were subsequently transferred onto a SiO$_{2}$/Si substrate that had been pre-cleaned with isopropanol and acetone in an ultrasonic bath. We have used a hBN/polypropylene carbonate (PPC)/polydimethylsiloxane (PDMS) stamp to pick up the exfoliated MoS$_2$ flakes. For constructing twisted homobilayers, we have used the tear-and-stack technique \cite{kim2016van}, which requires only a single 2D flake. In the tear-and-stack method, the stamp with hBN flake is aligned to cover roughly half of the TMD monolayer. After contacting the flake, the stamp is then heated, cooled, and lifted. The part of the monolayer in contact with the hBN adheres to it, while the other half remains on the substrate, causing the flake to tear when lifting the stamp. The sample stage (with the remaining half of the monolayer) is then rotated by the desired twist angle. Finally, the hBN (already carrying half of the flake) is realigned with the other half and picks it up, forming a twisted homobilayer with the targeted rotation angle. For our sample, a twist angle of $\sim2^\circ$ was chosen, resulting in a moiré potential with a wavelength of approximately 9 nm. Finally, graphene is picked up with the hBN in similar fashion, and deposited on the edge of the twisted bilayer to allow for electric contact. To achieve a clean top surface in the final heterostructure, we employ a flipping transfer method \cite{cui2017low, zeng2019high, lupke2020proximity} to place our stack of 2D materials onto a new substrate, whilst ensuring the last picked up layer ends up on top without ever contacting the polymer. For our substrate, mica is chosen over conventional SiO$_2$/Si substrates due to its atomically flat surface and optical transparency. To make the sample conductive, we added an indium contact through the method of micro-soldering \cite{girit2007soldering}. This technique involves directly depositing a low-melting-point metal onto the 2D layer without the need for resist processing and thus is contamination free. 

\subsection*{Measurement setup}
We have used pc-AFM and c-AFM modes in an Agilent 5100 system operating in contact mode with an AD-28-SS boron-doped diamond tip, due to its high conductivity and sharpness, which benefit the resolution. In our setup, the tip remains electrically grounded, and instead a voltage bias is applied to the twisted TMD sample via the graphene/indium contact. A fiber optic cable is positioned beneath the sample stage to direct light onto the sample. This fiber is connected to a TLS120Xe monochromator, which allows illumination with a precisely tunable, narrow-band light source. This enables controlled excitation of the sample with a specific wavelength of light.

\subsection*{Data processing}

The photoresponse was calculated as described above, calibrating the photocurrent with the dark current and the power of the monochromator, see also Ref. \cite{son2015layer}. Inspection of individual photocurrent maps showed no evidence of tip changes. This observation is further supported by comparison of the dark current maps recorded before and after the measurement series, which revealed only minor intensity differences, attributable to gradual thermal drift in the cantilever set-point. Crucially, either dark current map (before or after) can be used to calculate the photoresponse without affecting the results presented in the main text. Since no scan-line artifacts or tip changes were detected, no scan-line correction was applied. For accurate calculation of the photoresponse, each photocurrent map had to be precisely aligned with the corresponding dark current map. The alignment involved three steps. The first step already took place during the measurements: an area containing a recognizable landmark (a defect) was selected, as an alignment marker for manual correction of thermal drift. The second step involved rigid image registration using template matching. Finally, to correct for non-linear warping effects (a result of, for example, the hysteresis of the piezo controller), an elastic registration was performed using SimpleElastic, which is part of the Insight Segmentation and Registration Toolkit (ITK). This method was found to align our current-maps to perfection without introducing artifacts. As a control, the photoresponse was calculated twice: once by subtracting the initial dark current map and once using the final dark current map. The resulting responses were nearly identical, confirming that the choice of reference did not bias the results. To reduce the impact of shot noise, the data was first subjected to 2D median filtering with a 3$\times$3 kernel. The maps were then down sampled by binning 4$\times$4 pixel blocks into single data points. Finally, to ensure consistent global trends between PPS and SPI, we detrended each spectrum. This detrending compensates for small variations of the Schottky barrier height with photon energy, see Extended Data Figs.~\ref{fig:extdatafig2}i, so that PPS and SPI exhibit the same overall spectral trend without altering relative peak positions. The PPS spectra shown in the manuscript are averaged spectra over typically tens of recorded spectra, in order to increase the signal to noise ratio.

\subsection*{Extraction of the Schottky barrier height}

We extract the Schottky barrier height $\phi_B$ by fitting, separately, the forward and reverse branches of the dark $I$-$V$ characteristics to the thermionic-emission (TE) model~\cite{allain2015electrical} with ideality factor $n$:
\begin{equation}
I(V,T) \;=\; I_0(T)\,\exp\!\left(\frac{qV}{n k_{\mathrm B}T}\right)\,
\Bigg[1-\exp\!\left(-\frac{qV}{k_{\mathrm B}T}\right)\Bigg],
\qquad
\label{eq:TE1}
\end{equation}
\begin{equation}
I_0(T) \;=\; A\,A^* T^2 \exp\!\left(-\frac{q\phi_B}{k_{\mathrm B}T}\right),
\label{eq:TE2}
\end{equation}
where $q$ is the elementary charge, $k_{\mathrm B}$ is Boltzmann's constant, $T$ the lattice temperature, $A$ the electrical contact area, and $A^*$ the effective Richardson constant. In the fits, we fix $A^* = 120~\mathrm{A\,cm^{-2}\,K^{-2}}$ (commonly used for MoS$_2$ contacts) and supply $A$ as a user input (a realistic range is $25$-$75~\mathrm{nm}^2$).

In the forward-bias regime, the square-bracket term in Eq.~\eqref{eq:TE1} is close to unity for $V \gtrsim 0.1$~V, so that
\begin{equation}
\ln |I| \;\approx\; b_0 + b_1 V,
\quad\text{with}\quad
b_1=\frac{q}{n k_{\mathrm B}T}, 
\qquad 
b_0=\ln \textit{I}_0 .
\label{eq:lin}
\end{equation}
From $b_1$, we obtain the ideality factor $n = q/(b_1 k_{\mathrm B}T)$; from $b_0$, we obtain $I_0=\exp(b_0)$ and then
\begin{equation}
\phi_B \;=\; \frac{k_{\mathrm B}T}{q}\,\Big[\ln\!\big(A\,A^* T^2\big) - \ln I_0\Big].
\label{eq:phi}
\end{equation}

For the forward branch, we select a voltage window $0.1~\mathrm{V} \le V \le 1.2~\mathrm{V}$, and for the reverse branch $-0.7~\mathrm{V} \le V \le -0.1~\mathrm{V}$. Outside these intervals, the curves deviate from the linear trend in the $\ln|I|$ vs.\ $V$ plot, indicating transport mechanisms other than TE. As a goodness-of-fit diagnostic, we monitor the coefficient of determination $R^2$, which is consistently close to 1 for the selected voltage windows. In our implementation the TE fit treats $A^*$ as fixed and does not explicitly correct for image-force barrier lowering or series resistance. Deviations from ideal TE are partially captured by the fitted $n$ (with $n=1$ corresponding to the ideal TE limit). Therefore, the reported $\phi_B$ corresponds to an effective barrier under our contact geometry and bias conditions, suitable for comparative mapping across the moir\'e supercell rather than an absolute, quantitative barrier height. Overall, the extracted effective barriers are nearly polarity independent and spatially homogeneous, see Extended Data Figs.~\ref{fig:extdatafig2}f-h,  with an average value of $\sim 0.31$~eV. Small local variations are only correlated with the presence of point defects and not the moir\'e structure. Moreover, upon optical excitation, the barrier height increases slightly as a function of photon energy, see Extended Data Figs.~\ref{fig:extdatafig2}i. However, this effect is spatially uniform and does not alter the homogeneity of the barrier landscape.

\subsection*{Theory}

\subsubsection*{The Wannier equation}
Wannier excitons in TMDs are composed of a hole at the valence-band maximum and an electron at the conduction-band minimum, which form a bound state of the form
\begin{equation}
    \ket{X_{\kk_e,\kk_h}} = \sum_{\kk_e, \kk_h} \Phi(\kk_e, \kk_h) c^\dagger_{\kk_e}v^{}_{\kk_h} \ket{0},
\end{equation}
where $\kk_e$ and $\kk_h$ are the electron and hole momentum, respectively. Moreover, $\Phi(\kk_e, \kk_h)$ is the excitonic wavefunction in momentum space and $c^\dagger_\kk$ and $v^{}_\kk$ are the electronic creation operator for the conduction band and annihilation operator for the valence band, respectively. The behavior of the exciton wavefunction is governed by the Wannier equation \cite{Wannier, 2006PQE....30..155K}:
\begin{equation}
    (\varepsilon_{\kk_e}^c-\varepsilon^v_{\kk_h})\Phi(\kk_e, \kk_h) - \sum_\bq V(\bq)\Phi(\kk_e+\bq, \kk_h+\bq) = E_{}\Phi(\kk_e, \kk_h).
\end{equation}

%E_{\kk_e,\kk_h}\Phi(\kk_e, \kk_h).

Here, $\varepsilon_{\kk}^c$ and $\varepsilon^v_{\kk}$ are the dispersion relations of the conduction and valence band, and $E_{\kk_e,\kk_h}$ is the total exciton energy. The term $V(\bq)$ describes the attractive potential between electron and hole due to Coulomb interaction. In a three-dimensional system, this would be a screened Coulomb potential. For mono- (or few-) layer systems, such as TMDs, a modified potential should be used \cite{Merkl2019, Brem2020}: 
\begin{align}
     V(\bq) &= \frac{e^2}{2\epsilon_0 A q \epsilon(q)},\\ 
    \text{monolayer}: \qquad \epsilon(q) &= \kappa_\text{T} \tanh \left\{\frac{1}{2} \left[\xi_\text{T} q d - \log \left( \frac{\kappa_\text{T} - \kappa_{0}}{\kappa_\text{T} + \kappa_{0}}\right) \right]\right\},\\
    \text{bilayer}: \qquad  \epsilon(q) &= \begin{cases}
  \kappa_0 \frac{g_q f_q}{\cosh(\xi_\text{T} q d/2) h_q}      & \text{intralayer}\\
  \kappa_0 g_q^2 f_q         & \text{interlayer} 
\end{cases}, \label{eq:bilayer}
\end{align}
with the functions:
\begin{align}
    f_q &= 1 + \left( \frac{\kappa_\text{T}}{\kappa_0} +  \frac{\kappa_0}{\kappa_\text{T}}\right) \tanh(\xi_\text{T} qd) + \tanh^2(\xi_\text{T} q d), \\
    h_q &= 1 + \frac{\kappa_0}{\kappa_\text{T}}\left[ \tanh(\xi_\text{T} q d) + \tanh(\xi_\text{T} q d/2)\right] + \tanh(\xi_\text{T} q d) \tanh(\xi_\text{T} q d/2),\\
    g_q &= \frac{\cosh(\xi_\text{T} q d)}{\cosh(\xi_\text{T} q d/2)\left[ 1 + \frac{\kappa_0}{\kappa_\text{T}} \tanh(\xi_\text{T} q d/2) \right]}.
\end{align}
Here, $e$ is the elementary charge quantum, $\epsilon_0$ is the vacuum permittivity, $A$ is the system surface area, $q = |\bq|$, and $d$ is the monolayer thickness\footnote{The interlayer separation does not enter Eq.~\eqref{eq:bilayer} because the dielectric environment of bilayer TMDs is modeled by two homogeneous dielectric slabs on top of each other.}. The coefficients $\kappa = \sqrt{\epsilon^\parallel \epsilon^\perp}$ and $\xi = \sqrt{\epsilon^\parallel /\epsilon^\perp}$ are the dielectric constant and in-plane polarizability, respectively, with $\epsilon^{\parallel/\perp}$ the in-plane and out-of-plane dielectric constants. The subscript T and $0$ imply TMD and background, respectively.

Assuming parabolic dispersion around the valence band maximum and conduction band minimum for the kinetic term, we obtain
\begin{equation}
    \varepsilon_{\kk_e}^c-\varepsilon^v_{\kk_h} = \frac{\hbar^2 k_e^2}{2m_e} + \frac{\hbar^2 k_h^2}{2m_h} + E_\text{gap} = \frac{\hbar^2 k^2}{2m_r} + \frac{\hbar^2 Q^2}{2M} + E_\text{gap},
\end{equation}
with $m_r^{-1} = m_e^{-1} + m_h^{-1}$ and $M=m_e + m_h$, and $E_\textbf{gap}$ the gap between the valence and conduction band. We also  introduced the relative ($k= |\kk|$) and center of mass ($Q = |\QQ|$) momenta, where
\begin{equation}
    \kk = \alpha \kk_h + \beta \kk_e, \qquad \QQ = \kk_e -\kk_h,
\end{equation}
with $\alpha = m_e/M$ and $\beta=m_h/M$. Separating the exciton wavefunction in relative coordinate and center-of-mass coordinates, $\Phi(\kk,\QQ) = \Psi(\kk) \xi(\QQ)$, yields
\begin{align}
    &E_{\QQ,\mu} = \frac{\hbar^2 Q^2}{2M} + E_\text{gap} + E^b_\mu, \label{eq:Energy}\\
    &\frac{\hbar^2 k^2}{2m_r} \Psi_\mu(\kk) - \sum_\bq V(\bq) \Psi_\mu(\kk+\bq) = E^b_\mu \Psi_\mu(\kk),
\end{align}
where $E^b_\mu$ is the exciton binding energy and we have introduced the exciton state index $\mu$. Furthermore, here we assume that there is only a single valence-band maximum and conduction-band minimum at which excitons can form. In the remainder of this text, we adopt a more general definition where we add subscripts $ij$ to $\alpha$, $\beta$, $M$ and $\Psi$. This is to indicate that they describe excitons with an electron in the conduction-band minimum of valley $\zeta_i$ in layer $\ell_i$, and a hole in the valence-band maximum of valley $\zeta_j$ in layer $\ell_j$.

\subsubsection*{Binding energies}
For \mos the relevant valleys are $K$ (and $K'$) and $\Gamma$ for holes, and $K$ and $\Lambda$ for electrons. Using the parameters in Extended Data Table~\ref{tab:params}, the bilayer Wannier equation is solved for different background dielectric constants. For our sample, we approximate\footnote{$\kappa_0$ should account for the dielectric constant of the combined background, i.e. the vacuum above and the substrate (hBN) below. Vacuum has a dielectric constant of $1$ while hBN has a dielectric constant of $4-5$.} $\kappa_0 \approx 2$. Upon solving the Wannier equation for the effective mass parameters (see Extended Data Table \ref{tab:params}) corresponding to combinations of these valleys, we obtain the exciton binding energies. Extended Data Figs. \ref{fig:binding}a and \ref{fig:binding}b shows the binding energies for the indirect $\Gamma-K$ and $\Gamma-\Lambda$ excitons as a function of background dielectric constant. Here, the blue and orange curves denote the $1s$ and $2s$ intralayer exciton binding energies, while the green curve corresponds to the $1s$ interlayer exciton binding energy. Additionally, we present the binding energies for the direct $K-K$ ($K'-K'$) excitons in Extended Data Fig. \ref{fig:binding}c. The parameters for which the binding energies are calculated are given in Table~\ref{tab:params}.

\subsubsection*{Photon Energies}
To compare to the photon energies obtained experimentally, displayed in the main text, we calculate the total exciton energy, given by Eq.~\ref{eq:Energy} with $\QQ=0$.
Furthermore, $E^\text{gap}$ is the (indirect) energy gap between the relevant conduction-band minimum and valence-band maximum. From Extended Data Fig.~\ref{fig:binding}a, we obtain the binding energy $E_{\Gamma K}\simeq -200$ meV for the interlayer exciton (see green line). Combining this with a band gap \footnote{For bilayer \mos, the valence band maximum is at $\Gamma$ such that the $\Gamma-K$ gap is the smallest exciton gap, see Fig.~1h of the main text \cite{cheiwchanchamnangij2012quasiparticle, jin2013direct}}.~$E^\text{gap}_{\Gamma K} \approx 1.6\text{--}1.7$ eV, yields exciton energies corresponding to those observed in the main text; $E\approx 1.65-0.2=1.45$ eV, which agrees with the experimentally obtained $E=1.46$~eV for the interlayer $\Gamma-K$ exciton. 

The $A$ and $B$ excitons are formed in the $K-K$ (or $K^\prime-K^\prime$) valley. As a consequence of the strong spin-orbit splitting of the bands, at the $K$ point a spin-down $X_A$ and spin-up $X_B$ exciton form. By time-reversal symmetry, the spin is reversed at the $K^\prime$ point, see Extended Data Fig.~\ref{fig:MoS2}a.

Solving the Wannier equation for the $K-K$ excitons yields an intralayer exciton binding energy of approximately $-300$ meV (see blue line in Extended Data Fig. \ref{fig:binding}c for $\kappa_0=2$). The bandgap for $A$ excitons is given by $E^\text{gap}_{KK,A}=E^\text{gap}_{KK} \approx 2.15 \eV$ and the bandgap for $B$ excitons is given by $E^\text{gap}_{KK,B}=E^\text{gap}_{KK} + (\Delta_\text{vb}+\Delta_\text{cb})=2.15+0.16+0.03 \approx 2.313 \phantom{\text{a}}\eV$. Here, $\Delta_\text{vb}$ and $\Delta_\text{cb}$ are the valence and conduction band splitting due to spin-orbit coupling. Thus, we obtain the total $1s$ exciton energies through Eq.~\eqref{eq:Energy} (with $\QQ = \mathbf{0}$),
\begin{equation*}
    E_{X_A^{1s}} \approx 1.85 ~\eV, \quad 
    E_{X_B^{1s}}  \approx 2.013 ~\eV.
\end{equation*}
These values for the $1s$ excitons are in good agreement with the experimentally obtained values, see e.g. Figs.~\ref{fig:fig1}i-k. In identical fashion, the $2s$ excitons yield the energies
\begin{equation}
    E_{X_A^{2s}} \approx 2.05 ~\eV, \quad E_{X_B^{2s}}\approx 2.213~\eV.
\end{equation}
The former coincides roughly with the $1s$ $X_B$ exciton, which is also observed in Ref.~\cite{Vaquero2020}. The latter matches the peaks around $2.2 ~\eV$ presented in Fig.~\ref{fig:fig2}b and Figs.~\ref{fig:fig2}c, implying that these correspond to $2s$ $B$ excitons. Additionally, the interlayer exciton $(iX)$ binding energy is approximately $-200$ meV. In rhombohedral (3R) stacking, interlayer hole tunneling at the K valleys is symmetry-forbidden, which suppresses the oscillator strength of $K-K$ interlayer excitons, and often requires an external field or disorder to become visible in optical spectra. Consistent with this, most spectra of our R-stacked twisted bilayer lack the strong interlayer-exciton absorption. At some locations, see Fig.~\ref{fig:fig2}b, possibly aided by local disorder, we observe a peak around $\sim 1.95$ eV, corresponding to this interlayer exciton, $E_{iX_A^{1s}}$.\\

Finally, the more spurious $I$ peaks in Fig.~1i could be attributed to intralayer $\Gamma-\Lambda$ excitons. In bilayer \mos the conduction band minimum at the $\Lambda$ point lies higher than at the $K$ point \cite{cheiwchanchamnangij2012quasiparticle}, such that there is a slightly bigger gap than for $\Gamma - K$, for example. The binding energy of the $1s$ intralayer $\Gamma-\Lambda$ exciton is approximately $-200$ meV. With an indirect band gap of $E^\text{gap}_{\Gamma \Lambda} \approx 1.8-1.9$ eV, this would yield the features indicated in Fig.~1i,k.

\subsubsection*{Exciton Hamiltonian}
The effect of the moiré superstructure of the twisted bilayer \mos  can be modeled through the Exciton Hamiltonian \cite{Brem2020, PereaCausin2025, Bremhyb}, given by

\begin{align}
H &= \sum_{ij,\QQ} E_{ij,\QQ}\, X_{ij,\QQ}^\dagger X_{ij,\QQ}^{}
   + \sum_{i\neq j,\, l,\, \QQ,\, \bq} \mathcal{T}^{c}_{il,jl}(\bq)\,
     X_{jl,\QQ+\bq}^\dagger X_{il,\QQ}^{} \notag\\
  &\quad - \sum_{i\neq j,\, l,\, \QQ,\, \bq} \mathcal{T}^{v}_{li,lj}(\bq)\,
     X_{lj,\QQ+\bq}^\dagger X_{li,\QQ}^{}, \label{eq:exH}
\end{align}

Here, the first term describes the `free' intralayer ($X_{00}$ and $X_{11}$) and interlayer ($X_{01}$ and $X_{10}$) excitons, subject to the quadratic dispersion
\begin{equation}
    E_{ij,\QQ} = \frac{\hbar^2 Q^2}{2M_{ij}} + E_{ij,\mathbf{0}},
\end{equation}
which is the energy of an exciton with a electron in layer $i$ and a hole in layer $j$. The second and third terms in Eq.~\eqref{eq:exH} describe the hopping of the conduction band electron and valence band hole, respectively, of interlayer excitons. Through these last terms, the excitons are subject to a periodic modulation. The hopping coefficients are given by
\begin{align}
    \mathcal{T}^c_{il,jl}(\bq) &= T^c_{ij} (\bq) \mathcal{F}_{il,jl}(\beta_{jl}\bq),\\
    \mathcal{T}^v_{li,lj}(\bq) &= T^v_{ij} (\bq) \mathcal{F}_{li,lj}(-\alpha_{jl}\bq),\\
    \mathcal{F}_{ij,nm}(\bq) & = \int d\kk \Psi^*_{ij}(\kk)\Psi_{nm}(\kk+\bq). \label{eq:F}
\end{align}
The tunneling matrix elements for the $K$-point tunneling read $T_{ij}^{c/v}(\bq) \approx \theta^{c/v}_0\sum_{n=0}^2 \delta_{\bq, C_3^n(\mathbf{K}_j-\mathbf{K}_i)}$, with $\theta_0^c=2.1  \text{ meV}$, $\theta_0^v=14.4 \text{ meV}$ \cite{WangInterlayer}, and $C^n_3$ is the rotation matrix over $2\pi/3$ radians, applied $n$ times.\\

The dispersion of the above Hamiltonian is most easily obtained from a zone-folding approach, in which the total exciton momentum $\QQ$ is restricted to the first moiré Brillouin zone (mBZ). The moiré potential induced by interlayer tunneling connects excitonic states at different $\QQ$ through reciprocal lattice vectors of the moiré superlattice, leading to hybridization between inter- and intralayer excitons and the emergence of moiré minibands. To this extent, we introduce the zone-folded operators
\begin{equation}
    F_{ij,s,\QQ} = X_{ij, \QQ +s_1\mathbf{g}_1 + s_2\mathbf{g}_2}, \qquad s =(s_1, s_2),
\end{equation}
where $\QQ$ is restricted to the mBZ spanned by $\mathbf{g}_1$ and $\mathbf{g}_2$. Upon transforming Eq.~\eqref{eq:exH} into the zone-folded basis, we obtain \cite{Brem2020, Bremhyb}

\begin{align}
H &= \sum_{ij,s,\QQ} \Tilde{E}_{ij,\,\QQ + s_1\mathbf{g}_1 + s_2\mathbf{g}_2}
      F_{ij,s,\QQ}^\dagger F_{ij,s,\QQ}^{} \notag\\
  &\quad + \sum_{iji'j',ss',\QQ} \left( \Tilde{\mathcal{T}}^c_{iji'j',ss'} - \Tilde{\mathcal{T}}^v_{iji'j',ss'} \right)
      F_{ij,s,\QQ}^\dagger F_{i'j',s',\QQ}^{}, \label{eq:Hfolded}
\end{align}

with
\begin{align}
    \Tilde{E}_{ij,\QQ} &= \frac{\hbar^2 \left[\QQ - (\mathbf{K}_i-\mathbf{K}_j)\right]^2}{2M_{ij}} + E_{ij,\mathbf{0}},\\
    \Tilde{\mathcal{T}}^c_{iji'j',ss'} &= \delta_{j,j'}(1-\delta_{i,i'}) \left[ \Theta^{c,0}_{iji'j'} \delta_{s_1^{}, s_1'} \delta_{s_2^{}, s_2'} + \Theta^{c,1}_{iji'j'} \delta_{s_1^{}, s_1'+i'-i} \delta_{s_2^{}, s_2'}\right. \notag\\
    &\quad + \left.\Theta^{c,2}_{iji'j'} \delta_{s_1^{}, s_1'+i'-i} \delta_{s_2^{}, s_2'+i'-i} \right],\\
    \Tilde{\mathcal{T}}^v_{iji'j',ss'} &= \delta_{i,i'}(1-\delta_{j,j'}) \left[ \Theta^{v,0}_{iji'j'} \delta_{s_1^{}, s_1'} \delta_{s_2^{}, s_2'} + \Theta^{v,1}_{iji'j'} \delta_{s_1^{}, s_1'+j'-j} \delta_{s_2^{}, s_2'} \right. \notag\\
    &\quad + \left. \Theta^{v,2}_{iji'j'} \delta_{s_1^{}, s_1'+j'-j} \delta_{s_2^{}, s_2'+j'-j} \right],\\
    \Theta^{c,n}_{iji'j'} &= \theta^c_0 \mathcal{F}_{iji'j'}[\beta_{i'j'} C_3^n(\mathbf{K_i}-\mathbf{K}_j)],\\
    \Theta^{v,n}_{iji'j'} &= \theta^v_0 \mathcal{F}_{iji'j'}[-\alpha_{i'j'} C_3^n(\mathbf{K_i}-\mathbf{K}_j)].
\end{align}
Equation~\eqref{eq:Hfolded} can now be diagonalized for a finite amount of moire subbands, for example by truncating $\lvert s_i \rvert \le 4$. This will result in a folded bandstructure for momenta $\QQ$ in the mBZ. The tunneling terms $\Tilde{\mathcal{T}}$ give rise to hybridization between intra- and inter-layer excitons and a modulation of the exciton wavefunctions with the moire wavelength. Before diagonalizing Eq.~\eqref{eq:Hfolded}, we first obtain an explicit expression for $\mathcal{F}$.

\subsubsection*{Analytical form of $\mathcal{F}_{ij,nm}(\bq)$}
The exciton wavefunction overlap, $\mathcal{F}_{ij,nm}(\bq)$, can be calculated numerically using the wavefunctions obtained from the Wannier equation. However, we can also approximate them sufficiently well with hydrogen wavefunctions \cite{chernikov}, i.e.
\begin{equation}
    \Psi_{ij}(\bq) = \frac{N_{ij}}{(1+a_{ij}^2q^2)^{3/2}},
\end{equation}
where $a_{ij}$ is the effective Bohr radius of the exciton with an electron at $i$ and a hole at $j$, and $N_{ij}$ is a normalization factor. The value of $a_{ij}$ can be obtained by fitting the hydrogen wavefunction on the numerically obtained exciton wavefunction from the Wannier equation, see Extended Data Fig. \ref{fig:wf}.

Recognizing that Eq.~\eqref{eq:F} is a convolution in momentum space, we can obtain it by performing a Fourier transform, taking a product, and transforming back, i.e.
\begin{equation}
    \mathcal{F}_{ij,nm}(\bq)=\text{FT}\left[ \Tilde{\Psi}^*_{ij}(\rr) \Tilde{\Psi}_{nm}(\rr)\right](\mathbf{q}),
\end{equation}
where $\Tilde{\Psi}_{ij}(\rr) \equiv \text{FT}^{-1}[\Psi_{ij}(\bq)](\rr) = (N_{ij}/a_{ij}^2) \text{exp}(-r/a_{ij})$. Consequently,
\begin{align}
    \mathcal{F}_{ij,nm}(\bq)&= \frac{N_{ij}N_{nm}}{a_{ij}^2 a_{nm}^2 } \text{FT}\left[\exp(-r/\lambda_{ij,nm}) \right](\mathbf{q}), \notag\\
    &=\frac{N_{ij}N_{nm}}{a_{ij}^2 a_{nm}^2 } \frac{\lambda_{ij,nm}}{(1+\lambda^2_{ij,nm}q^2)^{3/2}},
\end{align}
where we introduced $\lambda^{-1}_{ij,nm} = a_{ij}^{-1} + a_{nm}^{-1}$. Finally, we fix the normalization constants $N_{ij}$ by requiring $\mathcal{F}_{ij,ij}(\mathbf{0}) = 1$, such that $1 = \mathcal{F}_{ij,ij}(\mathbf{0}) = N_{ij}^2/(4 a_{ij}^2)$. Then, we obtain
\begin{equation}
    \mathcal{F}_{ij,nm}(\bq) =\frac{4 a_{ij} a_{nm}}{(a_{ij} + a_{nm})^2 } \frac{1}{(1+\lambda^2_{ij,nm}q^2)^{3/2}}.
\end{equation}

\subsubsection*{Exciton band structure}
Now, we diagonalize Eq.~\eqref{eq:Hfolded} as a function of $\QQ$ for a twist angle of $2^\circ$. Here, we focus on the excitons in the $K-K$ valley. The moire exciton bandstructure in the absence of interlayer tunneling is shown in Fig. \ref{fig:MEX}a. In this case, there is no moiré potential and the spectrum consists of folded parabolic bands. Figure \ref{fig:MEX}b shows the band structure when interlayer tunneling is present. Now, the intra- and interlayer excitons can hybridize, which leads to additional energy gaps and a different exciton spectrum. The excitons at the $\gamma$ point in the mBZ are bright (yet they may still have different oscillator strengths) \cite{Brem2020}. Plotting the excitonic wavefunction of the lowest band at $\QQ = \gamma$, we obtain Fig.~\ref{fig:MEX}c. Here, the origin corresponds to the $\text{R}^M_M$ registry of the twisted bilayer MoS$_2$. In this figure, a clear increase in exciton wavefunction magnitude can be observed at the $\text{R}^M_M$ points, which is periodic in the moiré latticevector of the twisted bilayer. This modulation is in good agreement with the experimentally obtained results, see e.g. Fig.~2 of the main text.

% \begin{figure}
%     \centering
%     \includegraphics[width=0.5\linewidth]{Figures/MoS2_moire_bilayer_interstate.pdf}
%     \caption{Interlayer state, unit cell has been indicated.}
%     \label{fig:enter-label}
% \end{figure}

\subsection*{Data and Code availability}

The Source Data underlying the figures of this study and the Code used for this study are available at https://doi.org/10.4121/6f789d27-4204-48d4-bfaf-a330b7d7950c

\subsection*{Acknowledgments}
   PB acknowledges funding from the Dutch Research Council (NWO, grant numbers: OCENW.M22.123 and NWO Vidi VI.Vidi.233.019). LJMW, SJHB, and PB acknowledge financial support from the European Research Council; funded by the European Union (ERC, Q-EDGE, 101162852). Views and opinions expressed are however those of the author(s) only and do not necessarily reflect those of the European Union or the European Research Council. Neither the European Union nor the granting authority can be held responsible for them. LE, KV and CMS acknowledge the research program “Materials for the Quantum Age” (QuMat) for financial support. This program (registration number 024.005.006) is part of the Gravitation program financed by the Dutch Ministry of Education, Culture and Science (OCW). KW and TT acknowledge support from the JSPS KAKENHI (Grant Numbers 21H05233 and 23H02052), the CREST (JPMJCR24A5), JST and World Premier International Research Center Initiative (WPI), MEXT, Japan.

\subsection*{Author contributions}
LJMW and JV fabricated the samples. LJMW performed the experiments on the twisted bilayer and analyzed the data. SJHB, KV and JDV performed the experiments and analyzed the data on the monolayer samples. LE, RA and CMS did the theory calculations. TT synthesized the hBN crystals. KW performed the optical property evaluation for quality control of the hBN crystals. CMS supervised the theory calculations. PB conceived the idea, supervised the experiments and wrote the paper with input from all the authors. All authors contributed to the interpretation of the experimental and theoretical data and reviewed the paper. 

\subsection*{Competing interests} The authors declare no competing interests.

% Extended Data Figures
\newpage
\subsection*{Extended Data}

\newcounter{extdatafig}
\setcounter{figure}{0}
\renewcommand{\figurename}{Extended Data Fig.}
\begin{figure}[h]
    \centering
    \includegraphics[width=\linewidth]{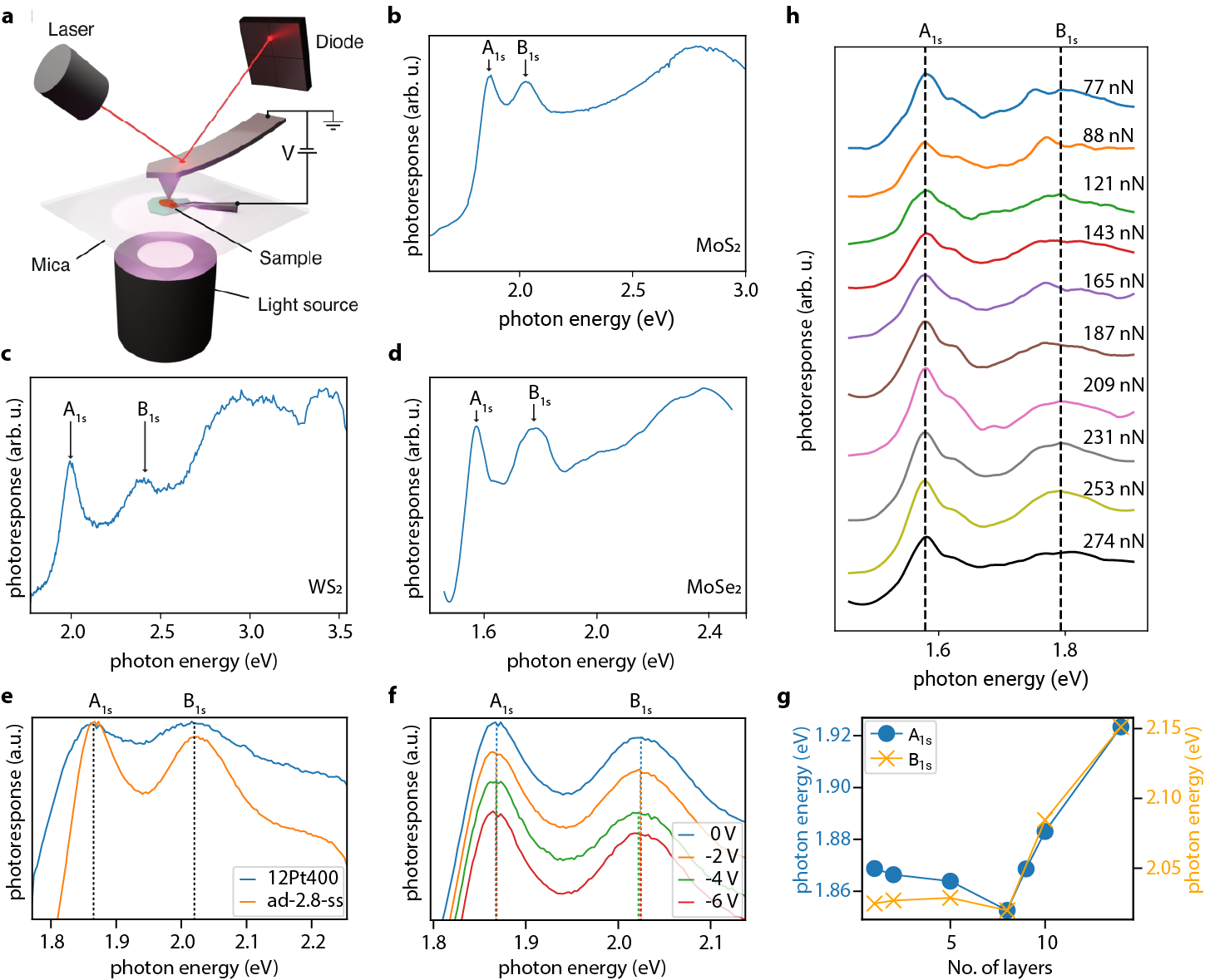}
    \caption{\textbf{pc-AFM setup and benchmarking.} \textbf{a,} Schematic of the pc-AFM experiment combining c-AFM with monochromatic back-illumination. \textbf{b,} \textbf{c,} and \textbf{d,} PPS spectra of monolayer MoS$_2$, WS$_2$ and MoSe$_2$, respectively. \textbf{e,} PPS spectra of monolayer MoS$_2$ acquired with two different tips (Pt (Rocky Mountain nano 12pt400b) vs doped diamond), showing the spectral convolution by the larger tip (Pt). \textbf{f,} PPS spectra of monolayer MoS$_2$ acquired for different sample biases, no noticeable differences are observed in the spectra. \textbf{g,} The A$_{1s}$/B$_{1s}$ exciton energy positions as a function of MoS$_2$ thickness. \textbf{h,} PPS spectra on monolayer MoSe$_2$ acquired as a function of the tip-induced force.}
    \label{fig:extdatafig1}
\end{figure}

\clearpage

\refstepcounter{extdatafig}

\begin{figure}
    \centering
    \includegraphics[width=\linewidth]{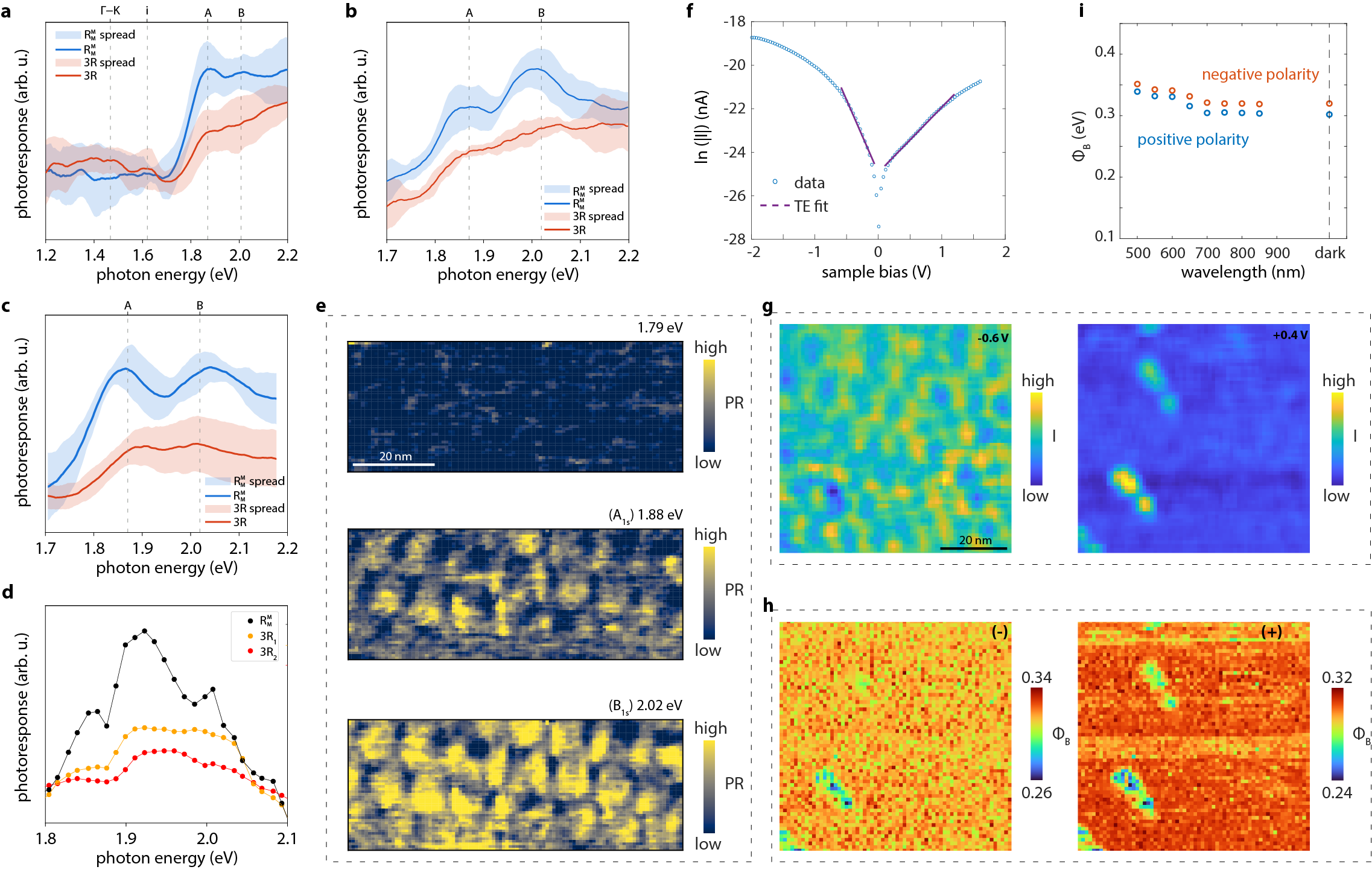}
   \caption{\textbf{Reproducibility of registry-resolved photoresponse.} \textbf{a-d,} Additional SPI data acquired at different sample locations showing the same registry-selective confinement of A$_{1s}$/B$_{1s}$ and indirect excitons as in the main text (blue marks spectra corresponding to the $R_{M}^{M}$ sites and red to 3R sites. \textbf{e,} PR(E) maps acquired away from the neutral excitons and at the energies of the A$_{1s}$ and B$_{1s}$ excitons, revealing the moir\'e modulation, in another location of the sample. \textbf{f,} Averaged ln(I)-V spectrum acquired on the twisted bilayer MoS$_2$ and fitted both polarities with the thermionic-emission model. \textbf{g,} Current images extracted from grid I-V measurements on the moir\'e. \textbf{h,} Maps of the calculated (effective) Schottky barrier heights for both the positive and negative polarity based on the thermionic emission model used to fit the curve in panel \textbf{f}, demonstrating a barrier homogeneity. \textbf{i,} Effective barrier height as a function of photon energy, showing a slight but spatially uniform increase.}
    
    \label{fig:extdatafig2}
\end{figure}

\clearpage

\refstepcounter{extdatafig}
\begin{table}[]
    \centering
      \captionsetup{name=Extended Data Table} % affects only this table

    \begin{tabular}{lc}
    \toprule
    Parameter & Value \\
    \midrule
    Lattice const. $a_0$ & \SI{0.317}{\nano\meter} \\
    Thickness $d$ & \SI{0.611}{\nano\meter} \\
    $\varepsilon_\parallel$  & 15.4 \\
    $\varepsilon_\perp$ & 6.1 \\
    $m_e^K$  & $0.43\,m_0$ \\
    $m_e^\Lambda$  & $0.64\,m_0$ \\
    $m_h^K$ & $0.54\,m_0$ \\
    $m_h^\Gamma$ & $2.5\,m_0$ \\
    $\Delta_\text{vb}$ & 140-160 meV \\
    $\Delta_\text{cb}$ & 3 meV \\
    \bottomrule
  \end{tabular}
  \captionof{table}{Material parameters for MoS$_2$, extracted from Ref.\cite{Kormanyos2015}.}
  \label{tab:params}
\end{table}

\clearpage
\refstepcounter{extdatafig}

\begin{figure}
    \centering
    \includegraphics[width=\linewidth]{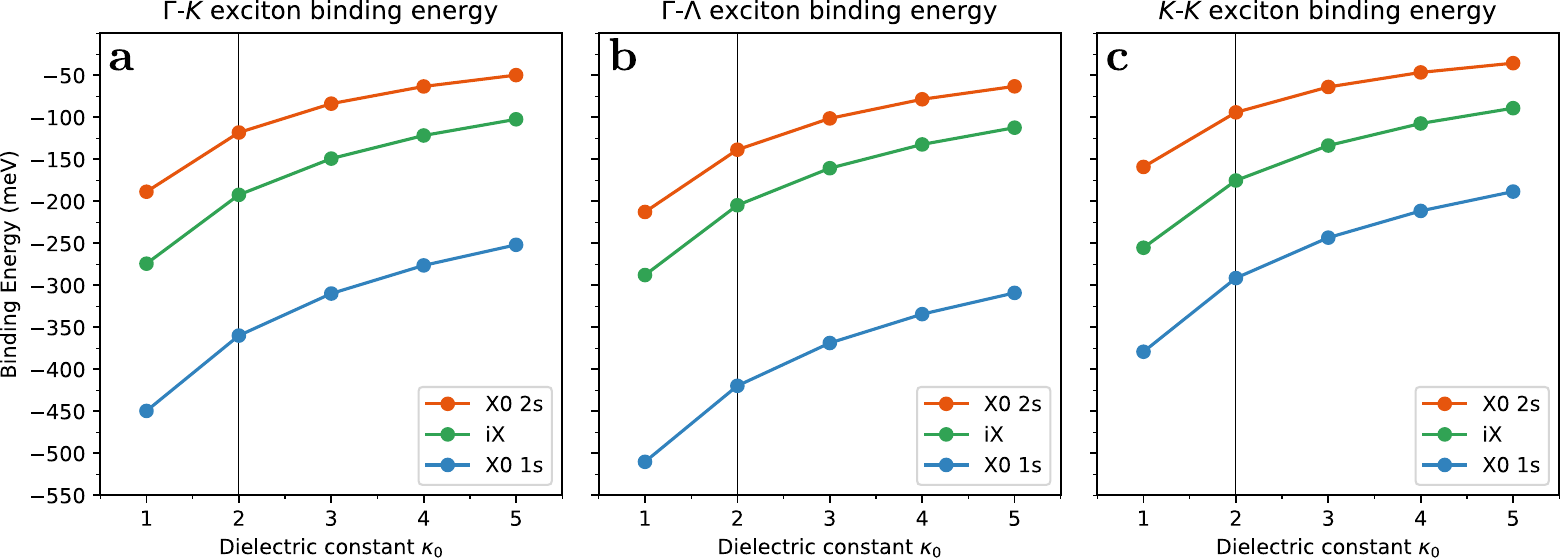}
    \caption{\textbf{Exciton binding energies}. Binding energies of the $1s$, $2s$ intralayer and $1s$ interlayer (iX) for \textbf{a,} indirect $\Gamma-K$; for \textbf{b,} indirect $\Gamma-\Lambda$, and \textbf{c,}  excitons as a function of background dielectric constant $\kappa_0$.}
    \label{fig:binding}
\end{figure}

\clearpage
\refstepcounter{extdatafig}

\begin{figure}[H]
    \centering
    \includegraphics[width=\linewidth]{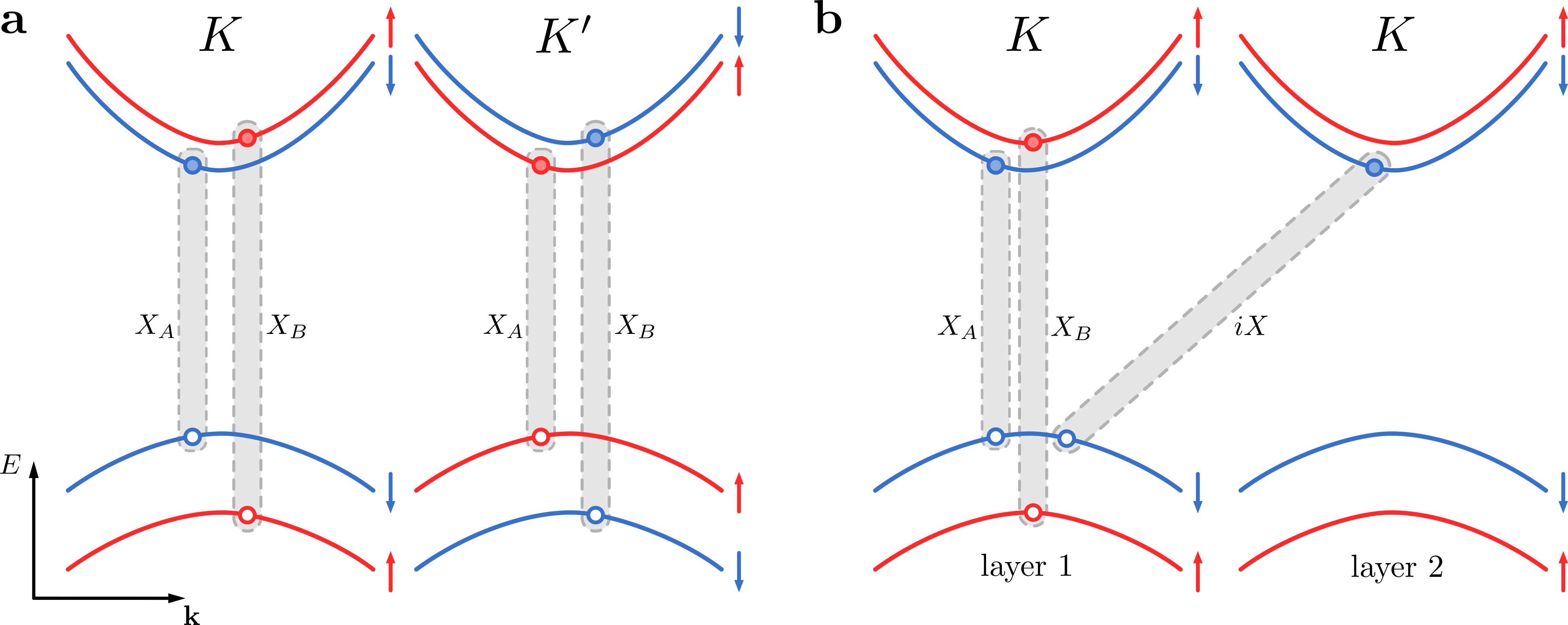}
    \caption{\textbf{Different types of excitons in MoS$_2$}. \textbf{a,} In monolayer MoS$_2$, $X_A$ and $X_B$ excitons form for spin down and spin up electrons/holes at the $K$ point, respectively. By time-reversal symmetry, the spin is reversed at the $K'$ point. Additionally, excitons may form between $K$ and $K'$ (not shown here). \textbf{b,} In (parallel stacked) bilayer MoS$_2$, additional interlayer excitons (iX) may form between the $K$ points of the different layers.}
    \label{fig:MoS2}
\end{figure}

\clearpage

\refstepcounter{extdatafig}

\begin{figure}[H]
    \centering
    \includegraphics[width=\linewidth]{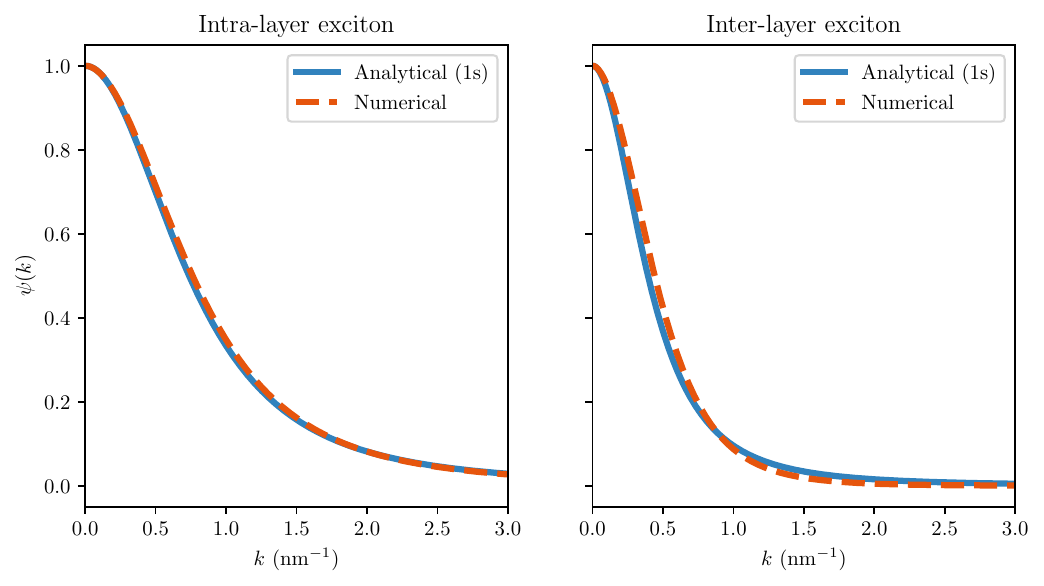}
    \caption{\textbf{interlayer and intralayer excitons}. Comparison between numerical and analytical (hydrogenic 1s) wavefunctions for intralayer (left) and interlayer (right) Wannier excitons in MoS$_2$.}
    \label{fig:wf}
\end{figure}
\refstepcounter{extdatafig}

\refstepcounter{extdatafig}

\refstepcounter{extdatafig}
\

% Reset to normal figure numbering for any subsequent figures
\newpage
\renewcommand{\thefigure}{\arabic{figure}}
\setcounter{figure}{0}

\end{document}